\documentclass[aps,prl,twocolumn,superscriptaddress]{revtex4-1}

\usepackage{graphicx}
\usepackage{amsmath}
\usepackage{amssymb}
\usepackage{graphicx}\usepackage{afterpage}
\usepackage{amsfonts}
\usepackage{amssymb}
\usepackage{textcomp}
\usepackage{color}
\usepackage{ulem}

\newcommand{\Ea}{\ensuremath{{\cal E}_1}}
\newcommand{\Eb}{\ensuremath{{\cal E}_2}}

\newcommand{\Er}{\ensuremath{{\cal E}_{\rm R}}}

\renewcommand{\vec}[1]{\bm{#1}}
\newcommand{\e}[0]{\text{e}}

\begin{document}
\title{Coherent coupling of individual quantum dots measured with phase-referenced two-dimensional spectroscopy: photon echo versus double quantum coherence}

\author{Valentin~Delmonte}
\affiliation{Univ. Grenoble Alpes, F-38000 Grenoble, France}
\affiliation{CNRS, Institut N\'{e}el, "Nanophysique et
semiconducteurs" group, F-38000 Grenoble, France}

\author{Judith~F.~Specht}
\affiliation{Institut f\"{u}r Theoretische Physik, Nichtlineare
Optik und Quantenelektronik, Technische Universit\"{a}t Berlin,
Hardenbergstrasse 36, 10623 Berlin, Germany}

\author{Tomasz~Jakubczyk}
\affiliation{Univ. Grenoble Alpes, F-38000 Grenoble, France}
\affiliation{CNRS, Institut N\'{e}el, "Nanophysique et
semiconducteurs" group, F-38000 Grenoble, France}

\author{Sven~H\"{o}fling}
\affiliation{Technische Physik, University of W\"{u}rzburg,
W\"{u}rzburg, Germany}

\author{Martin~Kamp}
\affiliation{Technische Physik, University of W\"{u}rzburg,
W\"{u}rzburg, Germany}

\author{Christian~Schneider}
\affiliation{Technische Physik, University of W\"{u}rzburg,
W\"{u}rzburg, Germany}

\author{Wolfgang~Langbein}
\affiliation{School of Physics and Astronomy, Cardiff University,
The Parade, Cardiff CF24 3AA, United Kingdom}

\author{Gilles~Nogues}
\affiliation{Univ. Grenoble Alpes, F-38000 Grenoble, France}
\affiliation{CNRS, Institut N\'{e}el, "Nanophysique et
semiconducteurs" group, F-38000 Grenoble, France}

\author{Marten~Richter}
\affiliation{Institut f\"{u}r Theoretische Physik, Nichtlineare
Optik und Quantenelektronik, Technische Universit\"{a}t Berlin,
Hardenbergstrasse 36, 10623 Berlin, Germany}

\author{Jacek~Kasprzak}
\email[]{jacek.kasprzak@neel.cnrs.fr}
\affiliation{Univ. Grenoble Alpes, F-38000 Grenoble, France}
\affiliation{CNRS, Institut
N\'{e}el, "Nanophysique et semiconducteurs" group, F-38000 Grenoble,
France}


\begin{abstract}

We employ two-dimensional (2D) coherent, nonlinear spectroscopy to
investigate couplings within individual InAs quantum dots (QD) and
QD molecules. Swapping pulse ordering in a two-beam sequence permits
to distinguish between rephasing and non-rephasing four-wave mixing
(FWM) configurations. We emphasize the non-rephasing case, allowing
to monitor two-photon coherence dynamics. Respective Fourier
transform yields a double quantum 2D FWM map, which is corroborated
with its single quantum counterpart, originating from the rephasing
sequence. We introduce referencing of the FWM phase with the one
carried by the driving pulses, overcoming the necessity of its
active-stabilization, as required in 2D spectroscopy. Combining
single and double quantum 2D FWM, provides a pertinent tool in
detecting and ascertaining coherent coupling mechanisms between
individual quantum systems, as exemplified experimentally.

\end{abstract}


\date{\today}

\maketitle

Nuclear magnetic resonance (NMR) spectroscopy conceived
phase-locked, multi-pulse techniques, yielding multi-dimensional
spectra by Fourier transforming temporal sequences into respective
frequency coordinates\,\cite{WuthrichNobelLecture,
VandersypenRMP05}. The possibility to spread the response of
biological or chemical molecules of high structural complexity,
especially proteins, across many axes enabled to assess their
spatial form and to understand inter-atomic interactions and
couplings. The idea to selectively address and evolve subsets of
transitions from congested spectra via multi-pulse toolbox, and then
projecting the results onto two-dimensional (2D) or
three-dimensional diagrams, is a far-reaching legacy of NMR. As
concern optical spectroscopy, achieving phase-stabilized sequences
of laser pulses is more challenging, owing to substantially shorter
optical cycles with respect to radio-frequency domain used in NMR.
Thus, it has been first accomplished in mid-NIR
regime\,\cite{HammJPCB98, MukamelARPC00, RehaultOE14} and recently
conveyed into NIR-VIS range\,\cite{BrixnerOL04, BrixnerJCP04,
BrixnerN05, ColliniN10, BaranczykAP14}, while being continuously
improved employing active pulse-shaping\,\cite{StoneScience09,
TurnerN10, Tollerud16} and phase-feedback
techniques\,\cite{BristowOE08, HelbingJOSAB11}.

At a juncture of coherent spectroscopy and condensed matter physics,
2D spectroscopy provided insight into dynamics and couplings of
many-body optical excitations in solids, in particular of excitons
in semiconductor quantum wells\,\cite{StoneScience09, TurnerN10,
MoodyPRL14} and novel 2D layered materials\,\cite{HaoNanoLett16}, as
well as in ensembles of quantum dots\,\cite{MoodyPRB13, MoodyPRB13a}
(QDs) or nanocrystals\,\cite{CassetteSmall16}. A principal tool in
these investigations is k-resolved four-wave mixing (FWM)
spectroscopy and its extensions probing multi-wave mixing
processes\,\cite{TurnerN10}. FWM microscopy of single QD
excitons\,\cite{LangbeinPRL05} was accomplished by phase-sensitive
optical heterodyning combined with interferometric detection,
efficiently subtracting resonant background and permitting co-linear
geometry of the excitation pulses. Recently, detection sensitivity
of intrinsically weak single QD FWM has been enhanced substantially
by using photonic nanostructures, improving the QD coupling with
external laser beams\,\cite{FrasNatPhot16, MermillodPRL16,
JakubczykACSPhot16}.

Here, we perform FWM spectroscopy of individual InAs QDs embedded in
a low-Q semiconductor
microcavity\,\cite{MaierOE14,FrasNatPhot16,MermillodOptica16}. We
point out two major advancements: Firstly, we demonstrate 2D FWM
constructed from two-photon coherences --- known as double quantum
2D FWM\,\cite{KimACR09, KaraiskajPRL10, TurnerPRB11} --- driven on
\emph{individual} transitions, specifically QD exciton-biexciton
systems (GXB)\,\cite{FingerhutAP13}. Secondly, we introduce
referencing of the FWM phase, offering convienient alternative for
its active-stabilization, which is widely believed to be required in
2D spectroscopy. Using the one-quantum and two-quantum spectroscopy,
we have measured single QDs and a QD molecule. A comparison of the
spectra signatures to theory allowed us to identify the nature of
the internal coupling mechanism in the QD molecule system. Our work
shows that the combined single and double quantum 2D spectroscopy is
a powerful tool to reveal and understand coherent coupling and
excitation transfer mechanisms - an interdisciplinary issue spanning
from biology and photo-chemistry, to quantum engineering. The
results are especially pertinent for the latter area, as we open new
avenues of research in quantum control of optically active
nanoscopic two-level and few-level systems in solids.

To acquire the FWM spectra\,\cite{FrasNatPhot16}, we use a pair of
100\,fs laser pulses: $\Ea$ and $\Eb$, with a variable delay
$\tau_{12}$, positive for $\Ea$ leading. They are frequency shifted
by $\Omega_1$ and $\Omega_2$, respectively, using acousto-optic
deflectors. FWM heterodyne beat with a reference field $\Er$ is
retrieved at $2\Omega_2-\Omega_1$ frequency, carrying the lowest
order response $\Ea^{\star}\Eb\Eb$ (where $\star$ denotes complex
conjugate) and also higher orders with the same phase evolution. The
signal is spectrally dispersed using a spectrometer, detected with a
CCD camera and retrieved in amplitude and phase by applying spectral
interferometry.

\begin{figure}[t]
\includegraphics[width=1.04\columnwidth]{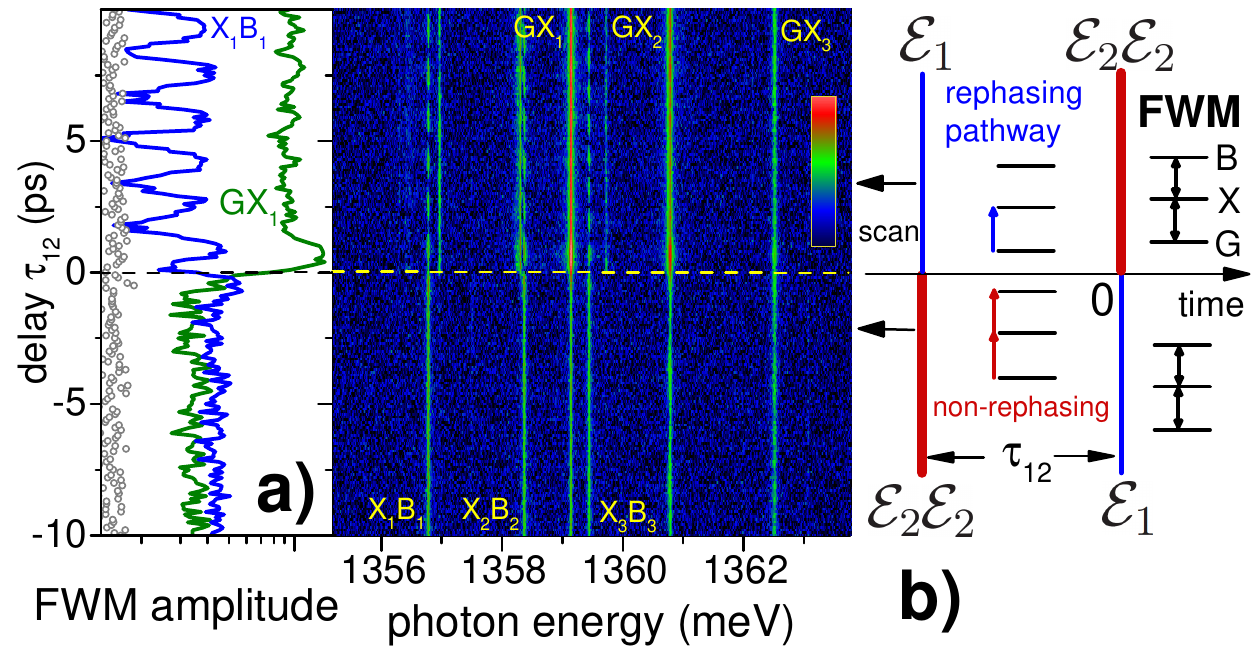}
\caption{{\bf Rephasing and non-rephasing pathways in two-beam
four-wave mixing.} (a)\,Typical measurement of the FWM amplitude as
a function of $\tau_{12}$ on a few InAs QDs embedded in a low-Q
microcavity. Impinging $\Ea$, $\Eb$ intensities of $(150,\,600)\,$nW
correspond to pulse areas of around $(0.4\pi,\,0.8\pi)$,
significantly beyond the $\chi^{(3)}$ limit. Positive (negative)
delays $\tau_{12}$ corresponds to rephasing (non-rephasing) FWM
pathways, as depicted in b). \label{fig:Fig1}}
\end{figure}

As shown in Fig.\,\ref{fig:Fig1}\,b, in two-beam FWM, the first
pulse $\Ea$ induces coherence, which evolves during $\tau_{12}$, to
be then converted into FWM by the second pulse $\Eb$, arriving a few
picoseconds after $\Er$. The lowest electronic excitations of a
neutral QD can be cast into three categories of states: a ground
state ($G$), single excitons ($X$) and two-exciton states, known as
biexcitons ($B$). $GX$ transitions are addressed by one-photon
coherence driven by $\Ea$, which is converted to FWM of $GX$ and
$XB$ by a density grating $\Ea^{\star}\Eb$ on $G$ and
$X$\,\cite{KasprzakNJP13, MermillodOptica16}. Inverting temporal
ordering of the two light pulses, $GB$ transition can be inspected
by a two-photon coherence induced by $\Eb$, transformed into FWM of
both transitions at the arrival of $\Ea$\,\cite{KasprzakNJP13,
MermillodOptica16}. The simple three-level system of
Fig.\,\ref{fig:Fig1}\,b illustrates the case of a neutral QD driven
along one of its polarization axes. For a single two-level system,
like a QD trion, FWM can be only created for $\tau_{12}>0$ from
one-photon coherence induced by $\Ea$, since the trion system cannot
be doubly excited within the employed spectral bandwidth. In fact,
two transitions in Fig.\,\ref{fig:Fig1}\,a show strictly no signal
for $\tau_{12}<0$ and are attributed to trion transitions. Therein,
we also recognize pairs of exciton-biexcitons, labeled as:
$GX_1-X_1B_1$, $GX_2-X_2B_2$, $GX_3-X_3B_3$ occurring in three
distinct QDs. FWM exhibits a pronounced beating as a function of
$\tau_{12}>0$, with a period corresponding to $B$ binding energy,
which is induced beyond $\chi^{(3)}$ regime by high order
contributions propagating at the FWM
frequency\,\cite{LangbeinJOSAB01, MermillodOptica16}. Instead, for
$\tau_{12}<0$ FWM is equally created on $GX$ and $XB$ transitions,
with no beating.

\begin{figure}[t]
\includegraphics[width=1.02\columnwidth]{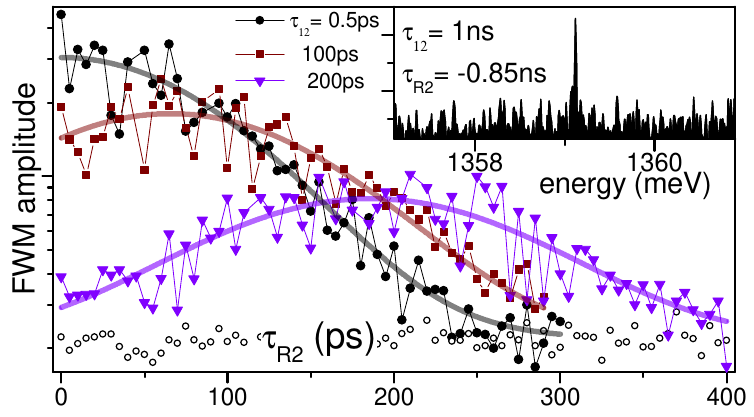}
\caption{{\bf Photon-echo formation on a single QD exciton measured
upon FWM rephasing pathway} $\tau_{R2}$ is scanned for different
values of $\tau_{12}$, as indicated. Temporal width of the echo
yields inhomogeneous broadening $\sigma$. \label{fig:Fig2}}
\end{figure}

Time-resolved FWM transient created upon the two pulse
configurations displays different characteristics. For
$\tau_{12}>0$, there is a phase-conjugation between $\Ea$ and FWM.
Owing to the rephasing, FWM of an inhomogeneously broadened system
has a Gaussian form, with a maximum at $t=\tau_{12}$ and temporal
width inversely proportional to the probed spectral inhomogeneous
broadening $\sigma$. Importantly, time-integrated amplitude of such
photon echo is not sensitive on $\sigma$, instead the homogeneous
broadening is probed through the $\tau_{12}$-dependence. At a level
of individual transitions, $\sigma$ is accumulated due to a residual
spectral wandering in time-averaged measurement\,\cite{PattonPRB06,
KasprzakNJP13, MermillodPRL16, JakubczykACSPhot16}. For $\sigma$ in
$\mu$eV range, which is a case even for high quality QD systems, the
echo width becomes comparable or larger than the temporal
sensitivity, given by the spectrometer resolution (here about
120\,ps). To demonstrate formation of such a broad
echo\,\cite{JakubczykACSPhot16}, we scan the delay $\tau_{R2}$,
between $\Er$ and $\Eb$, for three different $\tau_{12}$, as shown
in Fig.\,\ref{fig:Fig2}. The echo develops fully only for
$\tau_{12}=200\,$ps, from its width (FWHM)
$t_{\sigma}=\hbar/\sigma=(214\,\pm\,33)$\,ps we retrieve spectral
inhomogeneous broadening
$8\ln{(2)}\sigma=8\ln{(2)}\hbar/t_{\sigma}=(17\,\pm\,3)\,\mu$eV
(FWHM). Adjusting $\tau_{R2}$ permits to retrieve FWM significantly
beyond the temporal resolution of the setup, as shown in the inset
for $\tau_{12}=1\,$ns.

\begin{figure}[t]
\includegraphics[width=1.02\columnwidth]{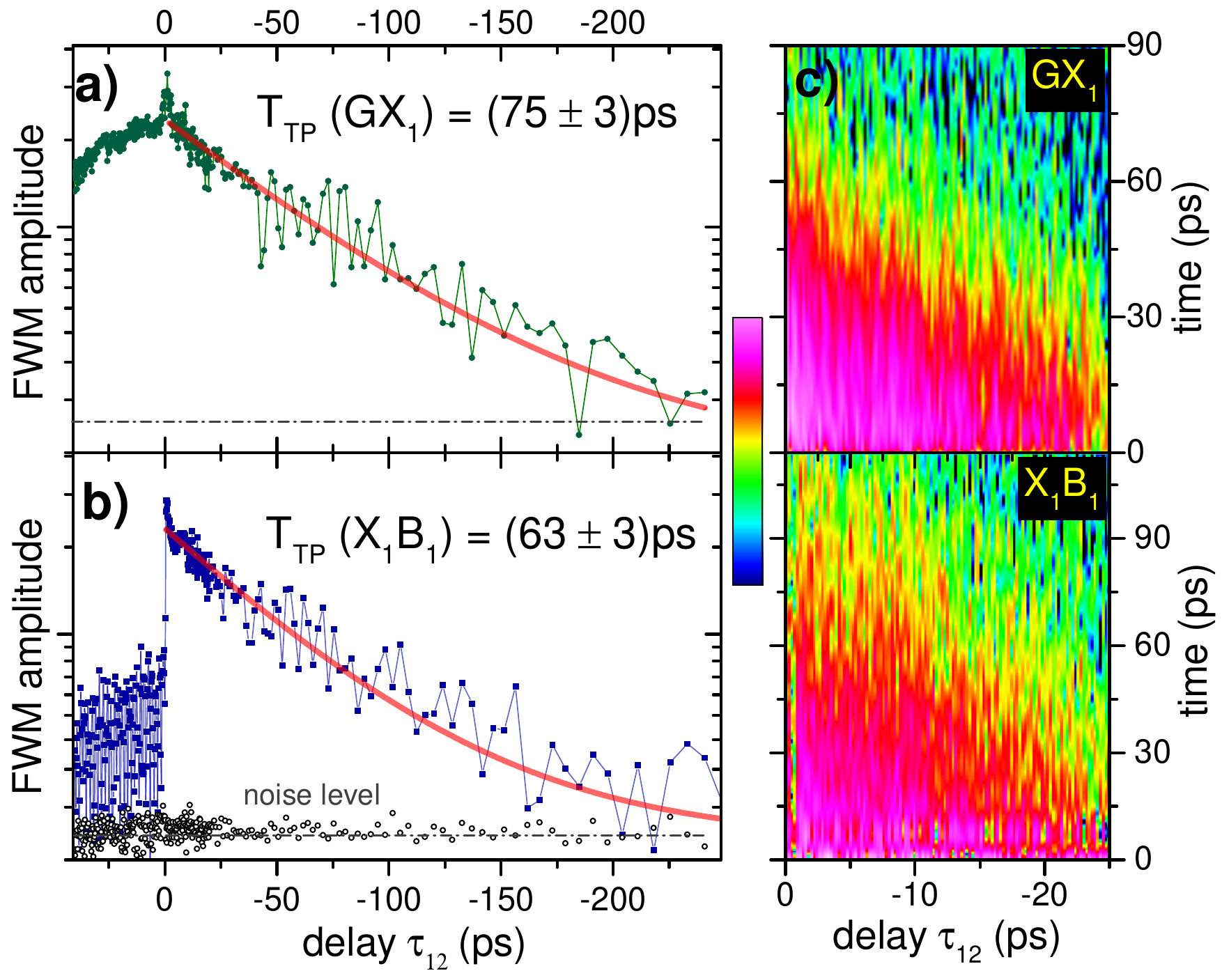}
\caption{{\bf Two-photon coherence dynamics and two-photon dephasing
of GX$_1$ and X$_1$B$_1$ measured at the non-rephasing FWM
configuration, $\tau_{12}<0$.} \label{fig:Fig3}}
\end{figure}

For $\tau_{12}<0$ there is no strict phase conjugation between
two-photon coherence and FWM, and therefore the photon echo is
absent. In Fig.\,\ref{fig:Fig3}\,c we show FWM$(t,\,\tau_{12}$) maps
measured on $GX_1$ and $X_1B_1$ transitions. As $\tau_{12}$ is
increased towards more negative values, FWM decay becomes more
pronounced, owing to a non-rephasing character of the signal. The
two-photon coherence dynamics of $GX_1$ and $X_1B_1$, i.e.
respective time-integrated FWM versus $\tau_{12}$, are presented in
Fig.\,\ref{fig:Fig3}\,a and b. From the exponential decay we
retrieve two-photon dephasing $T_{TP}(GX_1,
X_1B_1)=(75\,\pm\,3,\,65\,\pm\,3)\,$ps. Similar values of $T_{TP}$
are retrieved by analyzing two other $GX$-$XB$ pairs.

To illustrate couplings in the probed system of a few QDs, we
Fourier-transform FWM$(\omega_3,\,\tau_{12})$ sequences with respect
to the delay $\tau_{12}$. The experimental setup is encapsulated,
providing a passive stabilization of the phase during the
acquisition. However, the phase relationship between FWM measured
for subsequent $\tau_{12}$ is inevitably lost and can only be
achieved via active-stabilization\,\cite{BristowOE08,
HelbingJOSAB11}, which is not implemented here. The knowledge of the
FWM phase for subsequent delays $\tau_{12}$ is a precondition to
execute the Fourier transform yielding 2D FWM. In our previous
works\,\cite{KasprzakNPho11, MermillodOptica16}, we have
circumvented this issue by imposing a phase relationship onto the
data by choosing a separated transition in the spectral domain,
acting as a local oscillator, and setting its phase to zero for all
delays. We then applied this phase factor globally to the full
spectrum, adjusting all other frequencies versus $\tau_{12}$,
accordingly. Such transformation remains justified, as long as the
guiding transition to correct for, in particular exhibiting no
coherent coupling, is available in the spectrum. This generally
might not the case. To overcome this experimental limitation, we
have conceived a post-treatment protocol permitting to reference the
FWM phase, using auxiliary spectral interferences of $\Er$ with the
driving pulses, as explained in Supplementary Material.

\begin{figure}[t]
\includegraphics[width=1.02\columnwidth]{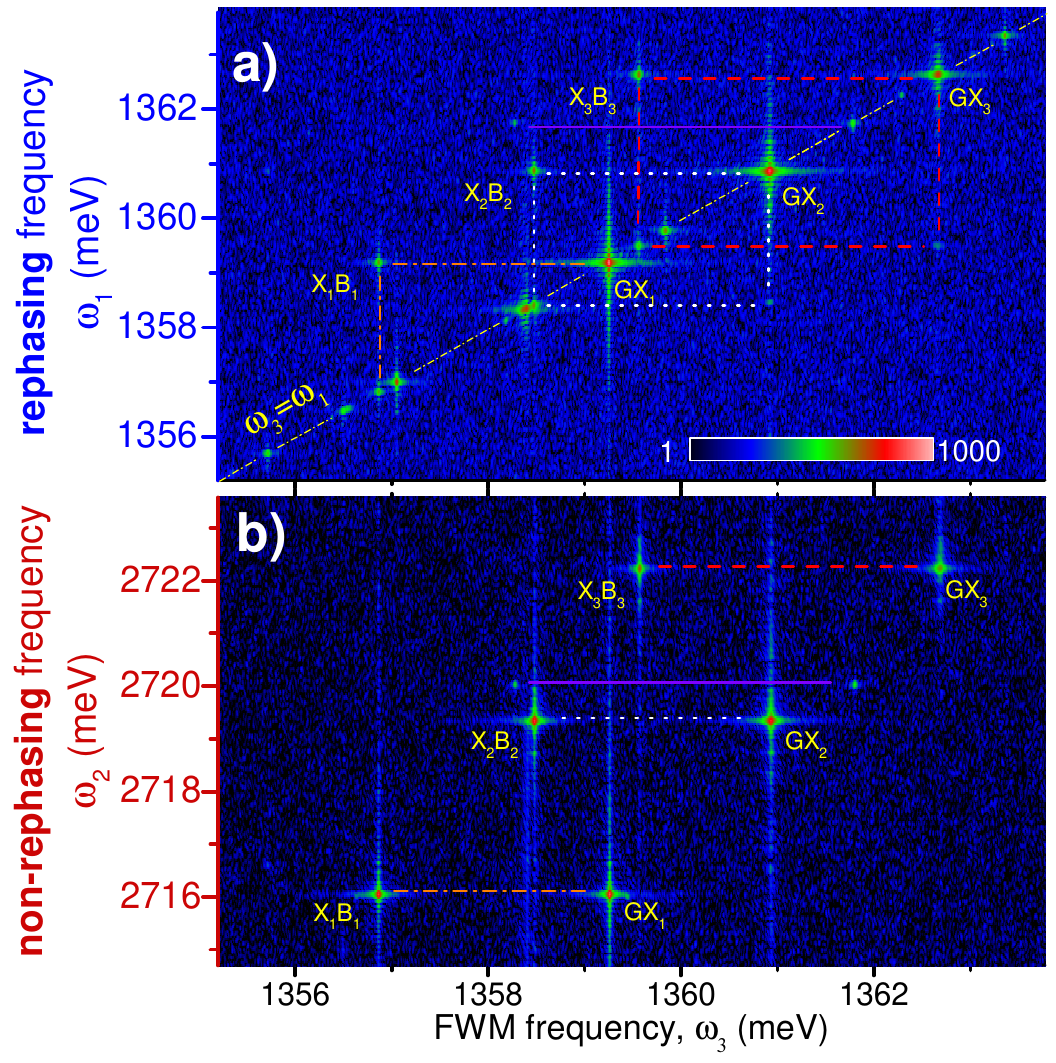}
\caption{{\bf Two-dimensional FWM spectroscopy of exciton complexes
in a few InAs QDs probed along the rephasing (a) and non-rephasing
(b) pathways.} Four exciton-biexciton systems in different QDs are
indicated by dash-dotted, dotted, dashed and solid lines,
respectively. \label{fig:Fig4}}
\end{figure}

In Fig.\,\ref{fig:Fig4} we present 2D FWM obtained from the set of
QDs highlighted in Fig.\,\ref{fig:Fig1}. For $\tau_{12}>0$, FWM
generated by all resonances driven by $\Ea$, forms a diagonal in the
resulting 2D spectrum. This includes, single trions and neutral
excitons, but also biexcitons - the latter can directly be driven by
$\Ea$ beyond the $\chi^{(3)}$ limit\,\cite{MermillodOptica16}, as
applied here (an example in the $\chi^{(3)}$ limit is provided in
the Supplementary Fig.\,S6). The biexcitons are off-diagonally
shifted by their respective binding energies of a few meV, and form
square-like features in 2D FWM under strong excitation, i.e. close
to $(\pi/2,\,\pi)$ area of $(\Ea,\,\Eb)$ pulses. 2D FWM resulting
from $\tau_{12}<0$ is shown in Fig.\,\ref{fig:Fig4}\,b. FWM
originates from a corresponding two-photon resonance driven by
$\Eb$. Here, the two-photon energy corresponds to the sum GX and XB
transition energies. In such non-rephasing 2D FWM, we retrieve the
response of GXB systems, whereas exciton complexes without doubly
excited states within the excitation bandwidth, such as singly
charged QDs, do not contribute.

\begin{figure}
\includegraphics[width=1.02\columnwidth]{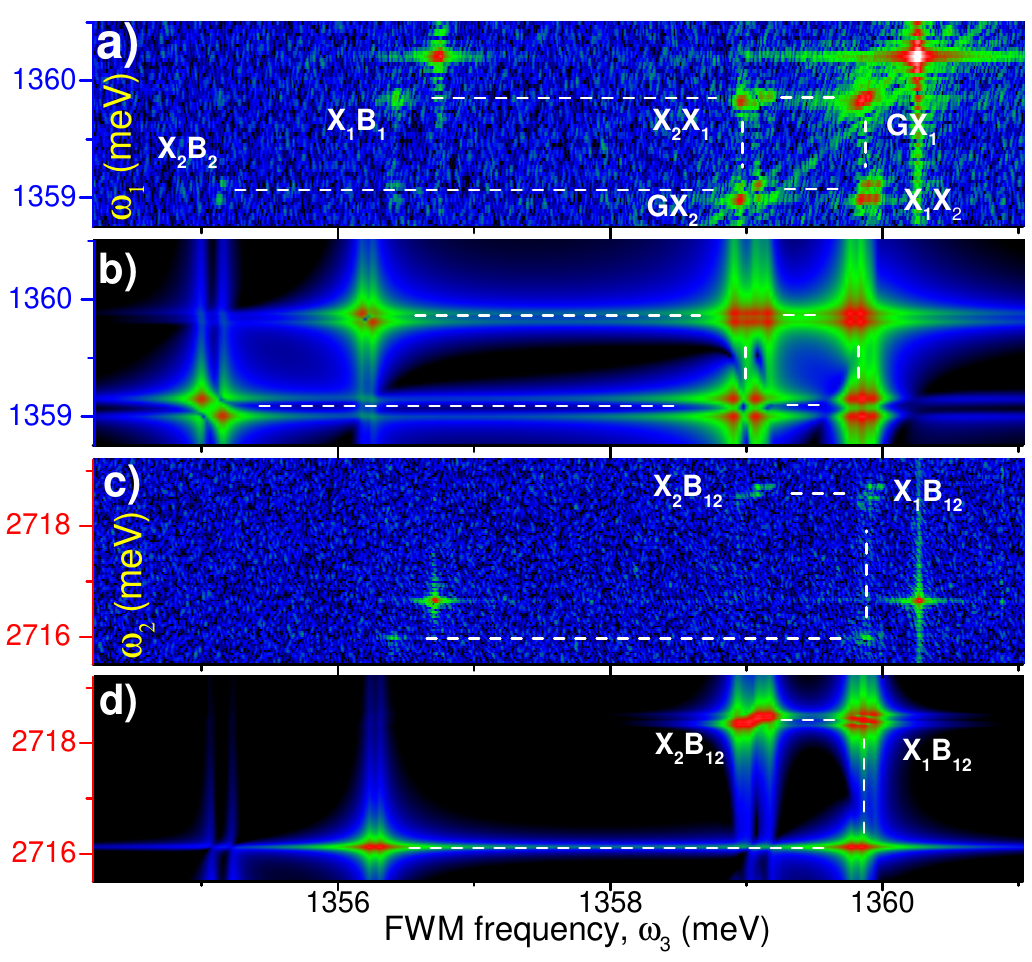}
\caption{{\bf Quantum dot molecule, consisting of two
electrostatically coupled InAs QDs, observed in single and double
quantum 2D FWM.} Measured rephasing (a) and non-rephasing (c) 2D FWM
spectra revealing coherent couplings between two QDs. Corresponding
simulations are shown in (b) and (d) with the parameters as listed
in the Supplementary Material. The signatures belonging to this QD
molecule are marked by dashed lines. Additional exciton-biexciton
pair in (a) and (c) at $(1360.3,\,1356.8)\,$meV occurs in other QD,
not involved in the molecule formation, thus not included in the
calculated spectra.} \label{fig:Fig5}
\end{figure}

Fig.\,\ref{fig:Fig5}\,a and c show the measured rephasing and
non-rephasing 2D spectra recorded at another position at the sample.
In the following, we focus on the two QDs that show up as
transitions $GX_1$ and $GX_2$ on the diagonal of the rephasing
spectrum with resonance energies $E_1$=1359.7\,meV and
$E_2$=1358.95\,meV - via hyperspectral
imaging\,\cite{KasprzakNPho11} these are found to be within
0.5$\,\mu$m vicinity (see Supplementary Fig.\,S7). The peak pattern
highlighted by the dashed lines differs from the signatures observed
in Fig.\,\ref{fig:Fig4} in two major respects: First, spin-orbit
coupling of the two circularly polarized excitons within each QD
leads to linearly polarized exciton eigenstates, where each QD is
described by a four-level system\,\cite{MermillodOptica16}. This
causes a splitting of each exciton resonance on the diagonal of the
rephasing spectrum into clusters of four peaks. Second, besides the
$X_1 B_1$ and $X_2 B_2$ peaks that are redshifted along the FWM axis
by the intradot biexciton binding energies $\Delta_1$=-3.3\,meV and
$\Delta_2$=-3.6\,meV, respectively, we observe two off-diagonal
cross peaks labeled $X_2 X_1$ and $X_1 X_2$ at the spectral
positions $(\omega_3=E_2; \omega_1=E_1)$ (upper cross peak) and
$(\omega_3=E_1; \omega_1=E_2)$ (lower cross peak). The appearance of
these cross peaks clearly indicates a coherent interdot coupling
between the two QDs: The electrostatic Coulomb coupling leads to an
energy renormalization of the interdot biexciton $B_{12}$ consisting
of one exciton in each QD. The biexciton shift lifts the symmetry of
the lower $GX_1$ ($GX_2$) and higher $X_1 B_{12}$ ($X_2 B_{12}$)
transitions such that the quantum pathways involving these
transitions do not destructively interfere anymore and cross peaks
show up\,\cite{KasprzakNPho11}. A level scheme of such a QD molecule
including all coupling-induced energy shifts is shown in
Supplementary Fig.\,1. The electrostatic interaction $\Delta_{12}$
between two excitons located in two different QDs is small compared
to the intradot biexciton binding energies. In fact,
spectrally-resolved FWM amplitude, shown in the Supplementary
Fig.\,S7, reveals that it is only of the order of
$\Delta_{12}=90\,\mu$eV and it shifts the interdot biexciton towards
higher energies, showing up as blueshifted high-energy shoulders of
the exciton resonance peaks. This interpretation is supported by
calculations\,\cite{AbramaviciusChemRev09} (see Supplementary
Material) of the rephasing and non-rephasing 2D signals depicted in
Fig.\,\ref{fig:Fig5}\,b and d.

In the non-rephasing two-quantum spectrum, the coupling of the two
QDs manifests itself in a peak pair labeled $X_1 B_{12}$ and $X_2
B_{12}$ at the interaction-shifted two-exciton transition $G B_{12}$
(energy $\omega_2=E_1+E_2+\Delta_{12}=$2718.74\,meV) with FWM
frequencies $\omega_3=E_1=$1359.7\,meV and
$\omega_3=E_2=$1358.95\,meV, respectively. Theoretical calculations
also suggest that exciton transfer processes between the two QDs
such as dipole-induced (F{\"o}rster) interaction and Dexter-type
coupling via wave-function overlap are
negligible\,\cite{SpechtPRB15}: First, these coupling types are
expected to be in the $\mu$eV range\,\cite{KasprzakNPho11} and
therefore difficult to detect at our spectrometer resolution of
25$\,\mu$eV. Second, they would lead to additional peaks for an
intradot biexciton in one QD after the first pulse has created a
single-exciton in the other QD. These peaks are not observed in the
spectra indicating that exciton transfer elements are negligible, as
elaborated in the Supplementary Material (Fig.\,S3).

An interesting feature about the observed QD molecule is that, in
contrast to the other isolated exciton-biexciton systems, the two
coupled QDs show a pronounced fine-structure splitting (FSS) of the
order of 60$\,\mu$eV and 140$\,\mu$eV, respectively. This is around
5 times higher than the FSS typically present in these
QDs\,\cite{MermillodOptica16}. Moreover, the FSS of the other
isolated exciton-biexciton systems in our sample (see also
Fig.\,\ref{fig:Fig4}) is not visible since the direction of the
linear excitation/reference polarization was chosen to be parallel
to the anisotropy axis. The observation of such a pronounced FSS
only for the resonances associated with the QD molecule therefore
suggests that the spatial proximity of the two coupled QDs altered
the local symmetry of the confinement, changing the magnitude of the
FSS and the polarization of the excitonic transitions.

In summary, we have implemented phase-referenced double quantum 2D
FWM spectroscopy of individual quantum systems. By merging it with
the single quantum counterpart, we have ascertained coherent
couplings between excitons, optical selection rules, the structure
of (bi-)exciton states, and coupling energies in single InAs QDs and
in a quantum dot molecule. This methodology is appealing to infer
electronic couplings and charge transfer in deterministically
defined QD molecules\,\cite{StinaffScience06, ArdeltPRL16} and
propagative coherence in photonic molecules\,\cite{KapfingerNC15}.
By merging it with recently developed multi-wave mixing
toolbox\,\cite{FrasNatPhot16}, it could be also used to visualize
and control polaritonic couplings in solid state cavity-quantum
electrodynamics\,\cite{AlbertNatComm13}.

\section*{acknowledgements} We gratefully acknowledge the financial
support by the European Research Council (ERC) Starting Grant PICSEN
(grant no. 306387). J.S. and M.R. acknowledge financial support by
SFB 787 and GRK 1558.


\newpage

\widetext

\begin{center}
{\bf \large SUPPLEMENTARY MATERIAL\\ Coherent coupling of individual
quantum dots measured with phase-referenced two-dimensional
spectroscopy: photon echo versus double quantum coherence\\}
Valentin~Delmonte, Judith~F.~Specht, Tomasz~Jakubczyk,
Sven~H\"{o}fling, Martin~Kamp, Christian~Schneider,
Wolfgang~Langbein, Gilles~Nogues, Marten~Richter, and Jacek~Kasprzak
\end{center}

\setcounter{figure}{0}

\renewcommand{\figurename}{Supplementary Figure}
\renewcommand{\thefigure}{S\arabic{figure}}

\section{Theory Notes} The model system used for the quantum dot (QD)
molecule is discussed and the third-order response function is
derived. The Liouville space pathways entering the rephasing photon
echo and non-rephasing double-quantum coherence signal are
illustrated using double-sided Feynman diagrams
\cite{AbramaviciusChemRev09}. Finally, the simulation parameters are
given and the calculated spectra are discussed with respect to the
coupling type of the QD molecule measured in Fig.\,5 of the main
manuscript.

\subsection{Model system of a quantum dot molecule}

\begin{figure}
    \centering
    \includegraphics[width=0.8\linewidth]{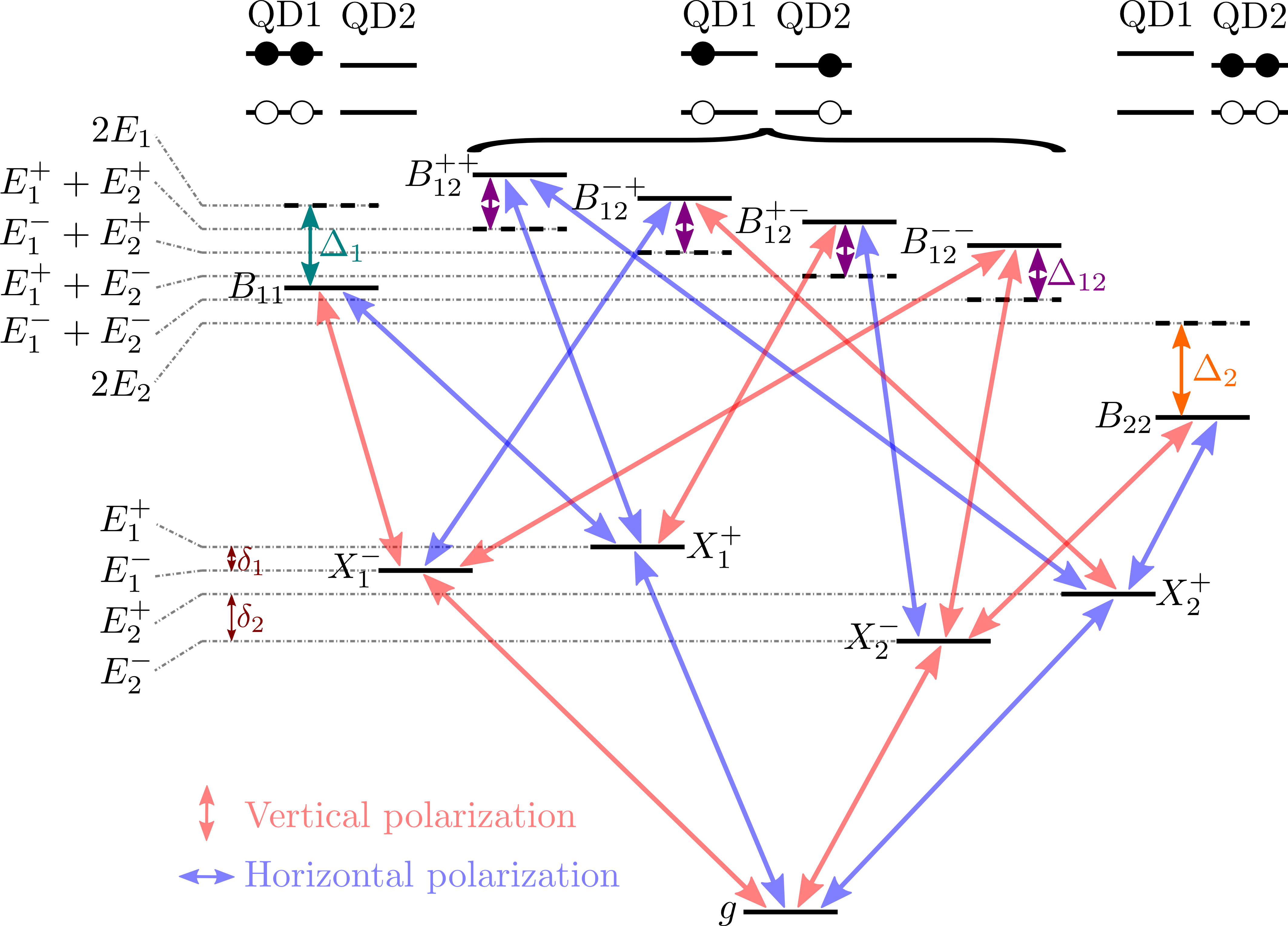}
    \caption{Schematic level scheme of a QD molecule consisting of two QDs. The exciton energies of the two QDs are $E_1$ and $E_2$. $\Delta_{12}$ is the electrostatic coupling of a doubly excited state, if one exciton is present in every QD. Each QD exhibits a fine-structure splitting $\delta_i$, which separates the QD excitons into an upper state $E_i^+ = E_i + \tfrac{\delta_i}{2}$ and a lower state $E_i^- = E_i - \tfrac{\delta_i}{2}$ ($i = 1,2$). For our QD molecule, the intradot biexciton binding energies $\Delta_1$ and $\Delta_2$ shift the biexciton states $B_{11}$ at energy $2 E_1$ and $B_{22}$ at energy $2 E_2$ towards lower energies, whereas the interdot coupling is repulsive, i.e., shifts the $B_{12}$ states towards higher energies.}
    \label{fig:SupplFig1}
\end{figure}

We will show that the signatures observed in the measured spectrum
of Fig.\,5 of the main text clearly indicate an intradot and
interdot coupling of excitons in a quantum dot (QD) molecule
consisting of two coupled QDs with resonance energies $E_1$ and
$E_2$, each with a different spin-orbit coupling $\delta_1$ and
$\delta_2$, respectively. The level scheme of such a QD molecule is
schematically shown in Fig.\,\ref{fig:SupplFig1}. There, the
electrostatic Coulomb coupling  shifts the energy of the doubly
excited states with respect to the single exciton energies of its
constituents. $\Delta_1$ ($\Delta_2$) denotes the intradot biexciton
shift within QD 1 (QD 2) and $\Delta_{12}$ represents the
electrostatic interdot coupling between an exciton in QD 1 and an
exciton in QD 2, forming a bound interdot biexciton $B_{12}$. The
optical selection rules are governed by the fine-structure splitting
(FSS): they originate from an exchange interaction between the two
circularly polarized excitons in one QD, resulting in a new linearly
polarized basis of excitonic eigenstates \cite{MermillodOptica16}.
Vertically polarized optical excitations are marked with red arrows,
horizontally polarized with blue arrows in
Fig.\,\ref{fig:SupplFig1}. Since the resonances of the two exciton
levels of each QD separated by the FSS are (more or less) equally
pronounced in the measured spectrum, we assume an angle of
$45^{\circ}$ between the direction of the linear
excitation/reference polarization and the fine-structure axis of the
two dots in our theoretical calculation. This way, all interaction
pathways shown in Fig.\,\ref{fig:SupplFig1} contribute to the
measured signal and a FSS will show up in the optical spectra.

\subsection{Calculation of the rephasing photon echo and
non-rephasing double-quantum coherence}

In the following, the rephasing one-quantum and non-rephasing
two-quantum signals are calculated following
Ref.\,[{\onlinecite{AbramaviciusChemRev09}}]. In a four-wave mixing
(FWM) experiment in the $\chi^{(3)}$ regime, the applied optical
field is composed of a sequence of three pulses with envelopes
$\mathcal{E}_j^{u_j}$ (where $\mathcal{E}_j^{-1} =
[\mathcal{E}_j^{+1}]^*$), frequencies $\tilde{\omega}_j$, and phases
$\varphi_j$. Each pulse $j$ is centered at time $\tilde{\tau}_j$:
\begin{equation}
    \vec{E} (\vec{r}, t) = \sum_{j=1}^3 \sum_{u_j=\pm 1} \mathcal{E}_j^{u_j} (\vec{r}, t - \tilde{\tau}_j) \e^{ i u_j (\varphi_j - \tilde{\omega}_j (t - \tilde{\tau}_j))}.
\end{equation}
The third-order response function links the polarization induced in
the system to the applied electric field:
\begin{align}
    \begin{split}
        P^{(3)}_\alpha (\vec{r},t)
        = & \int_{0}^\infty \int_{0}^\infty \int_{0}^\infty \text{d}t_3 \, \text{d}t_2 \, \text{d}t_1 \, \sum_{\beta,\gamma,\delta =1}^3 \; R^{(3)}_{\alpha \beta \gamma \delta } (t_3, t_2, t_1) \\
        & \times E_\beta (\vec{r},t - t_3) E_\gamma (\vec{r},t - t_3 - t_2) E_\delta (\vec{r},t - t_3 - t_2 - t_1).
    \end{split}
\end{align}
The heterodyned signal is a function of the delay times $\tau_{ij}
\equiv \tilde{\tau_j} - \tilde{\tau_i}$ between the pulses:
\begin{equation}
    S_{\Omega_{\text{s}}}^{(3)} (\tau_{3\text{s}}, \tau_{23}, \tau_{12}) = \int_{-\infty}^{+\infty} \text{d}t \, \vec{P}_{\Omega_\text{s}} (t) \cdot \mathcal{E}_s^* (t - \tilde{\tau}_s) \e^{i \tilde{\omega}_s (t - \tilde{\tau}_\text{s})},
\end{equation}
where $\mathcal{E}_\text{s} (t - \tilde{\tau}_\text{s})$ denotes the
local oscillator field envelope and $\Omega_\text{s} = u_1 \Omega_1
+ u_2 \Omega_2 + u_3 \Omega_3$ describes the specific phase
combination of the detected signal.

\begin{figure}
    \centering
    \includegraphics[width=0.7\linewidth]{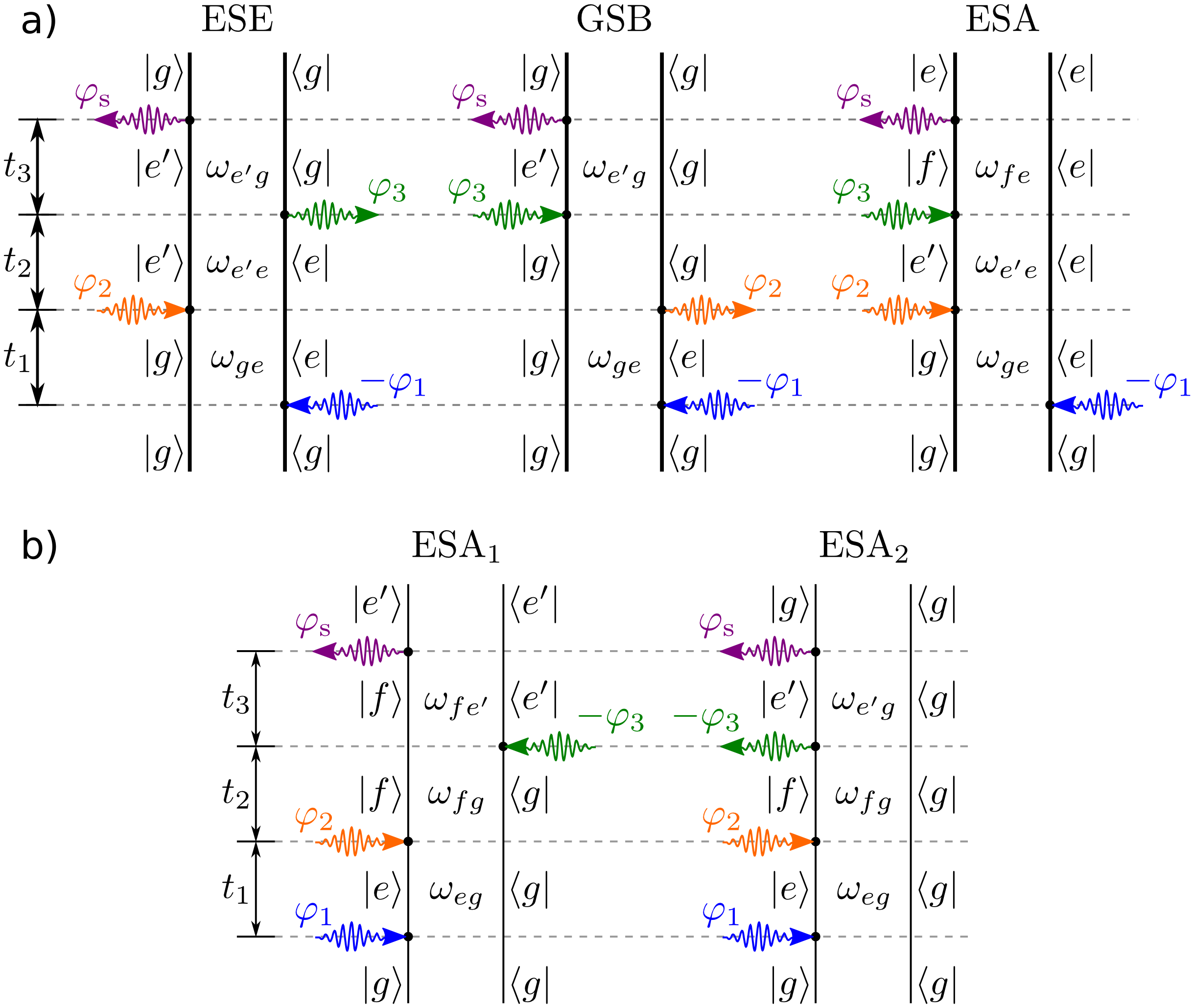}
    \caption{Double-sided Feynman diagrams illustrating the Liouville space pathways of the density matrix evolution. They enter the rephasing one-quantum signal $\Omega_\text{s} = - \Omega_1 + \Omega_2 + \Omega_3$ (a) and the non-rephasing two-quantum signal $\Omega_\text{s} = + \Omega_1 + \Omega_2 - \Omega_3$ (b).}
    \label{fig:SupplFig2}
\end{figure}

The rephasing photon-echo (PE) frequency combination is
$\Omega_\text{s} = \Omega_{\text{I}} = - \Omega_1 + \Omega_2 +
\Omega_3$. There are three types of excitation quantum pathways
contributing to the rephasing signal, denoted excited-state emission
(ESE), ground-state bleaching (GSB), and excited-state absorption
(ESA) and illustrated as double-sided Feynman diagrams in
Fig.\,\ref{fig:SupplFig2}(a). The ESE and GSB pathway involve only
the singly excited states $e$, whereas the ESA pathway includes also
the doubly excited states $f$. Note that in the absence of many-body
interactions, the two Liouville space pathways involving singly
excited states (lower transitions) are completely canceled out by
the pathway including the doubly excited states (upper transitions)
for off-diagonal resonances and therefore, all off-diagonal peaks
vanish\,\cite{DaiPRL12, KasprzakNPho11}.

The non-rephasing double-quantum coherence (DQC) frequency
combination is $\Omega_\text{s} = \Omega_{\text{III}} = + \Omega_1 +
\Omega_2 - \Omega_3$. The signal has two contributing ESA pathways,
cf. Fig.\,\ref{fig:SupplFig2}(b).

The rephasing PE one-quantum signal in frequency domain is
calculated by Fourier transforming the third-order
heterodyne-detected signal with respect to the time intervals $t_1$
and $t_3$ between the pulses. $t_2$ is fixed (In particular, in the
two-pulse FWM it is vanishing). The three contributions have the
form\,\cite{AbramaviciusChemRev09}
\begin{align}
    \begin{split}
        S_\text{ESE}^{(3)} (\omega_3, \tau_2, \omega_1) = \frac{(2 \pi)^4}{\hbar^3} \sum_{e,e^\prime} & \left( \vec{d}_{e^\prime g}^* \cdot \mathcal{E}_\text{s}^* (\omega_{e^\prime g} - \tilde{\omega}_s) \right) \left( \vec{d}_{e^\prime g} \cdot \mathcal{E}_2 (\omega_{e^\prime g} - \tilde{\omega}_2) \right) \e^{-i \xi_{e^\prime e} t_2} \\
        \times & \frac{\vec{d}_{e g} \cdot \mathcal{E}_3 (\omega_{e g} - \tilde{\omega}_3)}{\omega_3 - \xi_{e^\prime g}} \frac{\vec{d}_{eg}^* \cdot \mathcal{E}_1^* (\omega_{eg} - \tilde{\omega}_1)}{\omega_1 - \xi_{g e}},
    \end{split}\\
    \begin{split}
        S_\text{GSB}^{(3)} (\omega_3, \tau_2, \omega_1) = \frac{(2 \pi)^4}{\hbar^3} \sum_{e,e^\prime} & \left( \vec{d}_{e^\prime g}^* \cdot \mathcal{E}_\text{s}^* (\omega_{e^\prime g} - \tilde{\omega}_s) \right) \left( \vec{d}_{e g} \cdot \mathcal{E}_2 (\omega_{e g} - \tilde{\omega}_2)  \right) \\
        \times & \frac{\vec{d}_{e^\prime g} \cdot \mathcal{E}_3 (\omega_{e^\prime g} - \tilde{\omega}_3)}{\omega_3 - \xi_{e^\prime g}} \frac{\vec{d}_{eg}^* \cdot \mathcal{E}_1^* (\omega_{eg} - \tilde{\omega}_1)}{\omega_1 - \xi_{g e}},
    \end{split}\\
    \begin{split}
        S_\text{ESA}^{(3)} (\omega_3, \tau_2, \omega_1) = - \frac{(2 \pi)^4}{\hbar^3} \sum_{e,e^\prime,f} & \left( \vec{d}_{f e}^* \cdot \mathcal{E}_\text{s}^* (\omega_{f e} - \tilde{\omega}_s) \right) \left( \vec{d}_{e^\prime g} \cdot \mathcal{E}_2 (\omega_{e^\prime g} - \tilde{\omega}_2) \right) \e^{-i \xi_{e^\prime e} t_2} \\
        \times & \frac{\vec{d}_{f e^\prime} \cdot \mathcal{E}_3 (\omega_{f e^\prime} - \tilde{\omega}_3)}{\omega_3 - \xi_{f e}} \frac{\vec{d}_{eg}^* \cdot \mathcal{E}_1^* (\omega_{eg} - \tilde{\omega}_1)}{\omega_1 - \xi_{g e}},
    \end{split}
\end{align}
where we defined $\xi_{ab} \equiv \omega_{ab} - i \gamma_{ab}$.
$\vec{d}_{ab}$ denotes the dipole moment, $\omega_{ab}$ the
resonance energy and $\gamma_{ab}$ the homogeneous broadening of the
$b \rightarrow a$ transition. The total rephasing PE signal is the
sum of the three contributions:
\begin{equation}
    S_\text{PE}^{(3)} (\omega_3, t_2, \omega_1) = S_\text{ESE}^{(3)} (\omega_3, t_2, \omega_1) + S_\text{GSB}^{(3)} (\omega_3, t_2, \omega_1) + S_\text{ESA}^{(3)} (\omega_3, t_2, \omega_1).
\end{equation}

Similarly, the double-quantum signal is obtained by Fourier
transforming the signal function with respect to the delay times
$t_2$ and $t_3$. It is composed of the two ESA pathways:
\begin{equation}
    S_\text{DQC}^{(3)} (\omega_3, \omega_2, t_1) = S_{\text{ESA}_1}^{(3)} (\omega_3, \omega_2, t_1) + S_{\text{ESA}_2}^{(3)} (\omega_3, \omega_2, t_1)
\end{equation}
with
\begin{align}
    \begin{split}
        S_{\text{ESA}_1}^{(3)} (\omega_3, \omega_2, t_1) = - \frac{(2 \pi)^4}{\hbar^3} \sum_{e,e^\prime,f} & \left( \vec{d}_{fe^\prime}^* \cdot \mathcal{E}_\text{s}^* (\omega_{fe^\prime} - \tilde{\omega}_\text{s}) \right) \left( \vec{d}_{eg} \cdot \mathcal{E}_1 (\omega_{eg} - \tilde{\omega}_1) \right) \e^{-i \xi_{eg} t_1} \\
        \times & \frac{\vec{d}_{e^\prime g}^* \cdot \mathcal{E}_3^* (\omega_{e^\prime g} - \tilde{\omega}_3)}{\omega_3 - \xi_{fe^\prime}} \frac{\vec{d}_{fe} \cdot \mathcal{E}_2 (\omega_{fe} - \tilde{\omega}_2)}{\omega_2 - \xi_{fg}},
    \end{split} \\
    \begin{split}
        S_{\text{ESA}_2}^{(3)} (\omega_3, \omega_2, t_1) = \frac{(2 \pi)^4}{\hbar^3} \sum_{e,e^\prime,f} & \left( \vec{d}_{e^\prime g}^* \cdot \mathcal{E}_\text{s}^* (\omega_{e^\prime g} - \tilde{\omega}_\text{s}) \right) \left( \vec{d}_{eg} \cdot \mathcal{E}_1 (\omega_{eg} - \tilde{\omega}_1) \right) \e^{-i \xi_{eg} t_1} \\
        \times & \frac{\vec{d}_{f e^\prime}^* \cdot \mathcal{E}_3^* (\omega_{f e^\prime} - \tilde{\omega}_3)}{\omega_3 - \xi_{e^\prime g}} \frac{\vec{d}_{fe} \cdot \mathcal{E}_2 (\omega_{fe} - \tilde{\omega}_2)}{\omega_2 - \xi_{fg}}.
    \end{split}
\end{align}

These signal functions enable us to calculate 2D maps of the
rephasing and non-rephasing FWM pathways depending on the Fourier
transformed pulse delays.

\subsection{Model parameters} \label{sec:parameters}

The parameters used to calculate the rephasing and non-rephasing 2D
FWM spectra were chosen in agreement with the experimental data (see
Supplementary Tab.\,\ref{table:parameters}).

The homogeneous linewidth is typically in the order of few $\mu$eV,
which corresponds to a dephasing time of few hundred picoseconds
\cite{StockPRB11,OstapenkoPRB12}. Here, we choose $\gamma =
1/(500\,{\rm ps})$ \cite{BorriPRL01}. Moreover, the spectrometer
resolution of $\sim25\,\mu$eV was incorporated into the homogeneous
linewidth for a quick estimation of its effect. The transition
dipole moments of both QDs are chosen equally. Spectral wandering
induces an inhomogeneous broadening\,\cite{MermillodOptica16} of
$10\,\mu$eV. This is included in the calculations by averaging the
contributions to the signal functions for normally distributed
values of the system resonances.

\begin{table}[h!]
    \begin{tabular}{| c | c | c |}
        \hline
        &  QD 1 & QD2 \\
        \hline
        resonance energy & $E_1$ = 1359.7\,meV & $E_2$ = 1358.95\,meV \\
        \hline
        fine-structure splitting & $\delta_1$ = 60$\,\mu$eV & $\delta_2$ = 140$\,\mu$eV \\
        \hline
        biexciton binding energy &  $\Delta_1$ = -3.3\,meV & $\Delta_2$ = -3.6\,meV \\
        \hline
        interdot coupling & \multicolumn{2}{c |}{$\Delta_{12}$=+0.09\,meV} \\
        \hline
    \end{tabular}
    \caption{Model parameters used to calculate the optical response of the QD molecule composed of two coupled InAs QDs.}
    \label{table:parameters}
\end{table}

\subsection{Evaluation of the spectra}

Fig.\,5 and of the main text shows the calculated one-quantum
rephasing (a) and two-quantum non-rephasing (b) 2D spectra of the QD
molecule specified by the parameters given in
Sec.\,\ref{sec:parameters}. As mentioned before, Coulomb
interactions between the excited states cause that the doubly
excited states are energetically shifted compared to the sum of the
energies of the (isolated) single exciton constituents. This breaks
the symmetry between the $G \rightarrow X$ and $X \rightarrow B$
transitions and leads to off-diagonal peaks in the rephasing 2D
spectrum. In the measured spectrum in Fig.\,5\,(a) of the main text,
multiple off-diagonal peaks appear, indicating multiple exciton
couplings. The spin-orbit coupling splits the exciton resonances of
every QD, leading to four-peak clusters in the rephasing spectrum.
Moreover, the upper $X_2 X_1$ $(\omega_3 = E_2; \omega_1 = E_1)$ and
lower $X_1 X_2$ $(\omega_3 = E_1; \omega_1 = E_2)$ cross peaks form
a square with the two $GX_1$ and $GX_2$ resonances at energies $E_1$
= 1359.7\,meV and $E_2$ = 1358.95\,meV on the diagonal. This pattern
indicates an interdot exciton-exciton coupling between QD 1 and QD
2. (Without this coupling, these cross-peaks interfere destructively
and cancel.) The electrostatic interdot coupling $\Delta_{12}$ =
0.09\,meV is small compared to the linewidth and spectrometer
resolution. Thus, the corresponding interaction-shifted resonances
cannot be identified as separate peaks in the spectrum, but as
blueshifted high-energy shoulders of the exciton peaks, as
explicitly depicted in Supplementary Fig. X. Finally, the
off-diagonal $X_1 B_1$ and $X_2 B_2$ peaks at one-photon frequencies
$E_1$ and $E_2$ are redshifted along the FWM axis by the intradot
biexciton binding energies $\Delta_1$ = -3.3\,meV in QD 1 and
$\Delta_2$ =-3.6\,meV in QD 2. In the two-quantum spectrum, the
coupling of the two QDs leads to a peak pair labeled $X_2 B_{12}$
and $X_1 B_{12}$ at the interaction-shifted two-exciton transition
$G B_{12}$ at energy $\omega_2 = E_1 + E_2 + \Delta_{12}$ with FWM
frequencies $\omega_3 = E_1$ and $\omega_3 = E_2$, respectively.
Moreover, QD 1 (QD 2) forms a fine-structure split exciton-biexciton
complex at two-photon frequency $2 E_1 - \lvert \Delta_1 \rvert$ ($2
E_2 - \lvert \Delta_2 \rvert $). The corresponding peak pairs are
labeled $G X_i$ and $X_i B_i$ with $i \in 1,2$.

\begin{figure}
\includegraphics[width=0.9\linewidth]{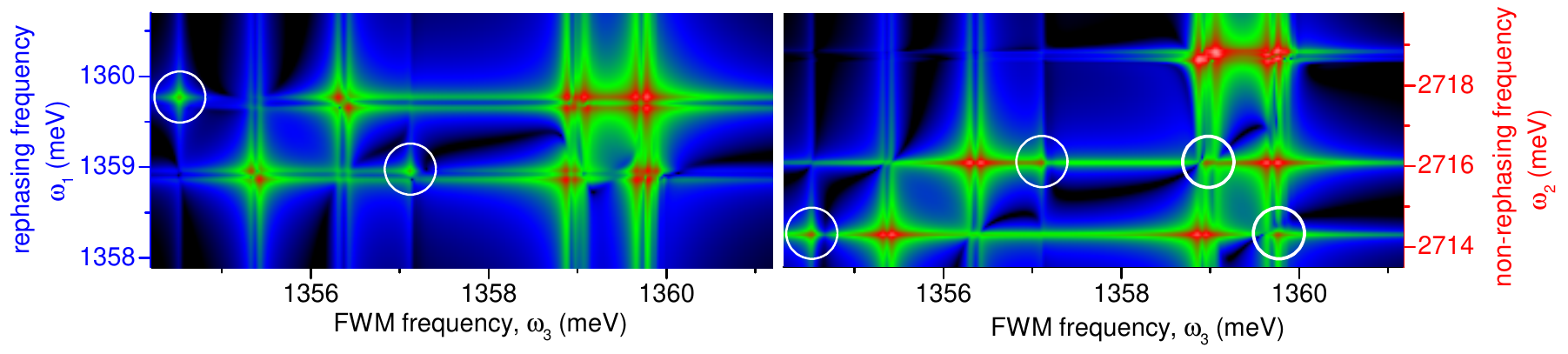}
\caption{Calculated rephasing one-quantum (a) and non-rephasing
two-quantum (b) 2D FWM spectra including F\"{o}rster coupling. The
additional peaks arising due to exciton transfer processes between
the QDs are marked by white circles\textcolor{red}{.}}
\label{fig:SupplFig3}
\end{figure}

A comparison of the measured and calculated spectra also shows that
exciton energy transfer processes between the two quantum dots such
as dipole-induced (F\"{o}rster) interaction and Dexter-type coupling
via wave-function overlap are negligible in the considered system:
First, the corresponding coupling strengths in the order of
$\mu$eV\,\cite{KasprzakNPho11} are too small to be detected for the
achieved spectrometer resolution. Second, the exciton transfer
between the two QDs would lead to additional peaks at spectral
positions where the second pulse creates a $GB$ coherence involving
an intradot biexciton in one QD before the third pulse induces a
$GX$ or $XB$ coherence involving a single exciton in the other QD.
To illustrate this, Suppl. Fig.\,\ref{fig:SupplFig3} shows the
calculated spectra including a (large) F\"{o}rster coupling of
0.1\,meV between the two QDs. The additional peaks arising due to
interdot exciton transfer are highlighted by white circles.

This way, the analysis of the measured and calculated spectra
enables us to identify the coupling type and strength, providing
evidence of a QD molecule consisting of two electrostatically
coupled QDs.

\section{\textbf{Auxiliary Experimental Results}}

\subsection{Implementation of the phase-referencing protocol}

In this section, we present the protocol which we established to
reference the phase of the FWM signal by using the excitation pulses
$\Ea$ and $\Eb$. For all measured delays $\tau_{12}$, before and
after acquiring FWM signal at $2\Omega_2-\Omega_1$ frequency, we
measure heterodyne spectral interferences at $\Omega_1$ and
$\Omega_2$, allowing to measure the phase of $\Ea$ and $\Eb$:
$\varphi_1$ and $\varphi_2$, by performing spectral interferometry.
Passive stabilization of the setup turns out to be sufficient to
maintain the phase stability for each $\tau_{12}$ during a typical
acquisition time of ten seconds, as required here. If however phase
drift is detected, the supplementary phase-reference data are
interleaved in between the FWM acquisition to monitor phase
variation and to correct for it.

\begin{figure}[t]
\includegraphics[width=\columnwidth]{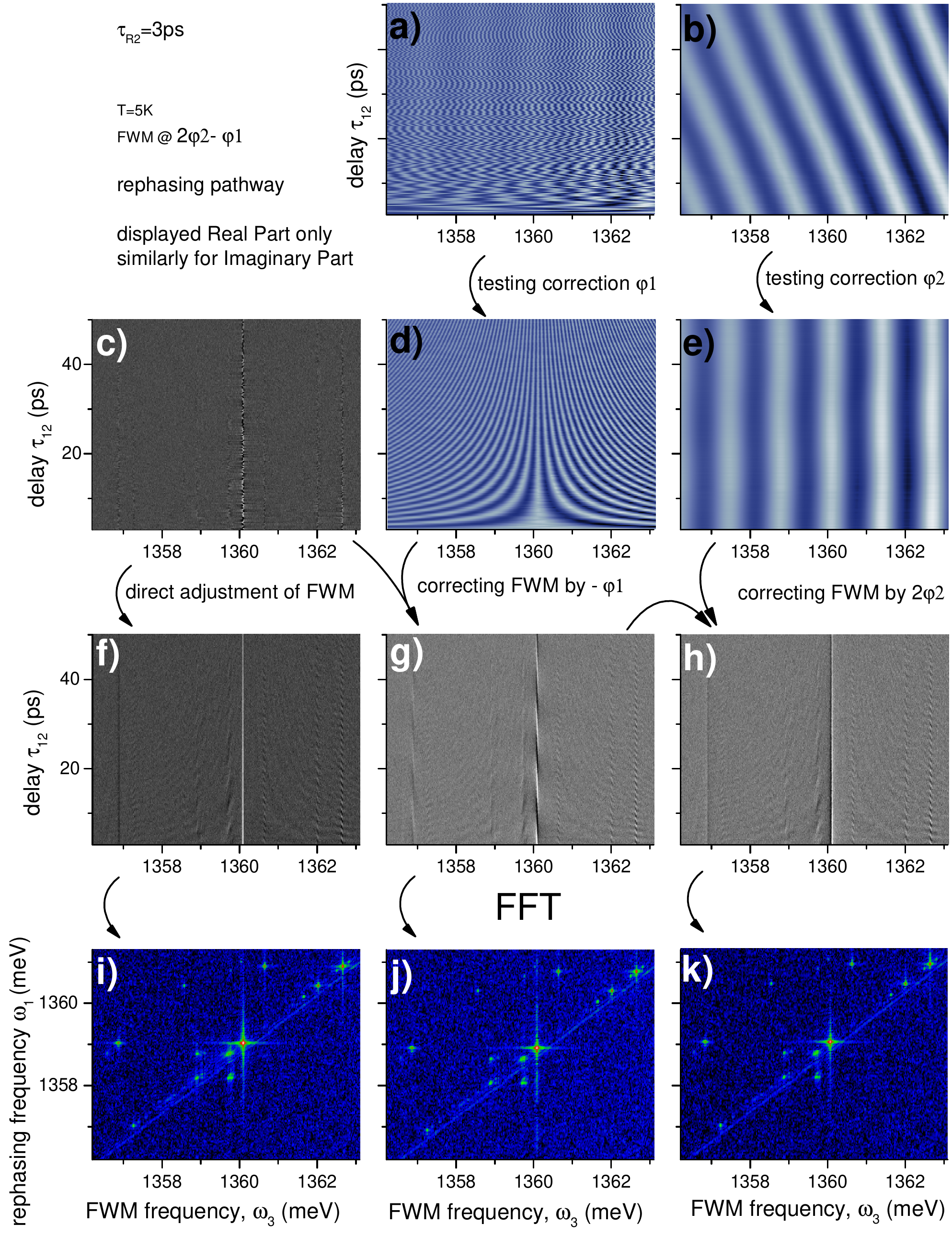}
\caption{{\bf Phase-referencing protocol. Rephasing pathway.
\label{fig:Figphaserefpos}}(a,\,b,\,c)\, Spectrally-resolved
heterodyne signal at $\Omega_1$, $\Omega_2$ and $2\Omega_2-\Omega_1$
respectively. (d,\,e,\,f)\,Phase-correction applied on (a,\,b,\,c),
respectively. (g,\,h)\,Phase-correcting FWM data (c), using (d,\,e),
respectively. (i,\,j,\,k)\,2D FWM retrieved from (f,\,g,\,h),
respectively. Only the real part of the data displayed. The
correction is analogously applied to the imaginary part, which is
not shown here.}
\end{figure}

In Fig.\,\ref{fig:Figphaserefpos}\,a and b we present the spectrally
resolved real part of the signal heterodyned at $\Omega_1$ and
$\Omega_2$, respectively. Here, $\tau_{R2}=3\,$ps and remains fixed
during the measurement, whereas $\tau_{12}$ (and thus also delay
between $\Ea$ and $\Er$, $\tau_{R1}$) is scanned. As a result,
$\varphi_1$ shows strong fluctuations, while $\varphi_2$ displays a
slope of only 2$\pi$ through the whole sequence.

In the next step, we test the performance of the phase correction
routine described in the main text. In panels d) and e) we see that
$\varphi_1$ and $\varphi_2$ can be aligned with respect to a chosen
frequency of around 1360.1\,meV.

We will now impose this externally determined correction onto the
FWM data shown c). Specifically, the FWM is corrected by phase
factors -$\varphi_1$ and $2\varphi_2$, which are retrieved from d)
and e), respectively. The real part of the FWM, phase-corrected
using such external referencing is shown in panel h). The FWM phase
is directly linked with the one of $\Ea$ and $\Eb$, which are now
measured and provide the reference frame at which we synchronize the
FWM. The accuracy of the phase-referencing is confirmed by formation
of narrow peaks in 2D FWM displayed in panel k). For comparison,
panel i) shows 2D FWM obtained after correcting the phase directly
on the FWM data.

\begin{figure}[t]
\includegraphics[width=\columnwidth]{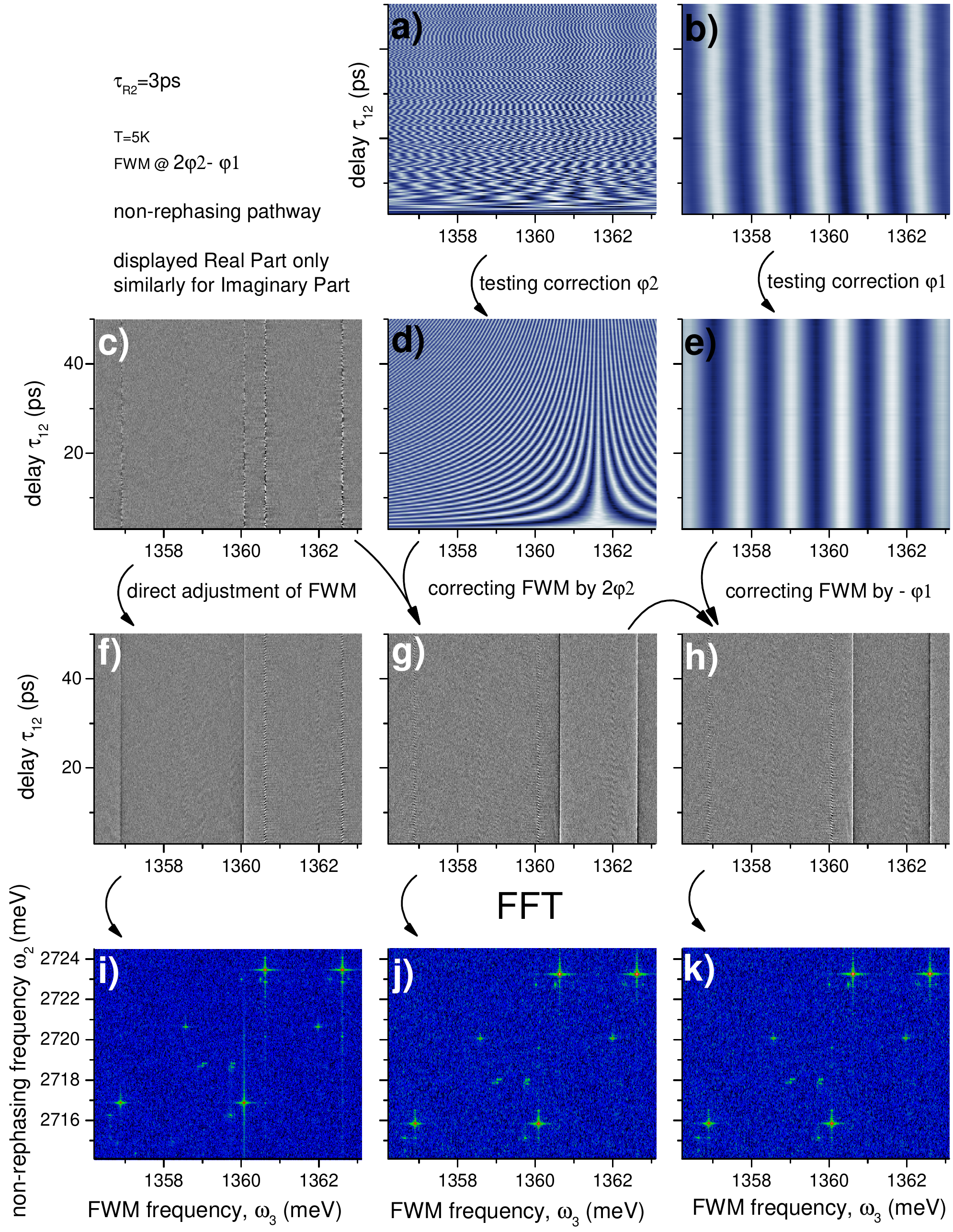}
\caption{{\bf Phase-referencing protocol. Non-rephasing pathway.
\label{fig:Figphaserefneg}}(a,\,b,\,c)\, Spectrally-resolved
heterodyne signal at $\Omega_2$, $\Omega_1$ and $2\Omega_2-\Omega_1$
respectively. (d,\,e,\,f)\,Phase-correction applied on (a,\,b,\,c),
respectively. (g,\,h)\,Phase-correcting FWM data (c), using (d,\,e),
respectively. (i,\,j,\,k)\,2D FWM retrieved from (f,\,g,\,h),
respectively. The correction is analogously applied to the imaginary
part, which is not shown here.}
\end{figure}

In Fig.\,\ref{fig:Figphaserefneg} we present phase-referenced 2D
FWM, obtained in the non-rephasing configuration, $\tau_{12}<0$.

\subsection{Single and double quantum 2D FWM of individual
quantum dots in a low excitation regime}

\begin{figure}[t]
\includegraphics[width=\columnwidth]{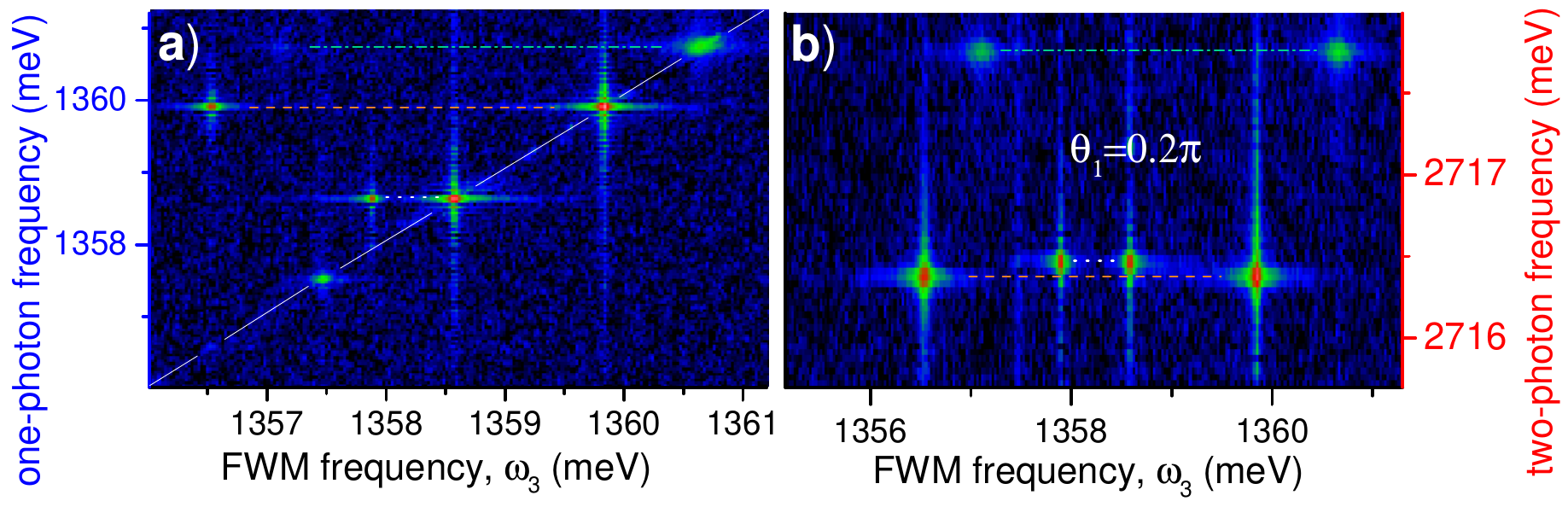}
\caption{{\bf 2D FWM spectroscopy excition-biexciton systems in
three distinct InAs QDs in a low excitation
regime.\,}(a)\,Rephasing, one-photon configuration. White line
depicts the diagonal $\omega_3=\omega_1$. (b)\,non-rephasing,
two-photon configuration. The applied pulse areas of $\Ea$ and $\Eb$
was $(\theta_1,\,\theta_2)\simeq(0.2, 0.6)\,\pi$, co-linear
polarization. The amplitude in logarithmic color scale over 2 orders
of magnitude. \label{fig:FigS6}}
\end{figure}

In Fig.\,\ref{fig:FigS6} we present single (a) and double quantum
(b) 2D FWM spectroscopy of individual QDs obtained under low
excitation $\chi^{(3)}$ regime. The spectra reveal exciton-biexciton
pairs, occurring in different QDs. In (a), the biexcitons are
shifted from the excitons --- placed at the diagonal (white line)
$\omega_3=\omega_1$ --- by their binding energies of -0.7\,meV
(white dotted line), -3.3\,meV (orange dashed line), -3.55\,meV
(green dash-dotted line). We note that, in contrast to the data
shown in Fig.\,4 a of the main manuscript, the signatures of
biexcitons do not show up on the diagonal. This means the ground
state-to-biexciton transitions are now not excited by $\Ea$,
confirming the low excitation regime.  By inspecting the
exciton-biexciton pair at (1360.67,\,1357.1)\,meV we observe that,
in the rephasing configuration, the FWM amplitude of the biexciton
is an order of magnitude weaker than the exciton's one. Instead, in
the non-rephasing situation the amplitude of both transitions is
equal. Such a suppression of biexciton transition in the rephasing
case is characteristic for transitions exhibiting a large
inhomogeneous broadening (here around 0.2\,meV, an order of
magnitude larger then typically observed in this sample), and has
been previously noticed in FWM of GaAs quantum
wells\,\cite{LangbeinPSSA02b}. Two other transitions, with an order
of magnitude lower inhomogeneous broadening, show a textbook
behavior with the exciton's amplitude twice stronger than (equal to)
the biexciton in the rephasing (non-rephasing)
configuration\,\cite{KasprzakNJP13, MermillodOptica16}.

\subsection{Experimental insight into a QD molecule: FWM amplitude and
phase, and hyperspectral imaging}

\begin{figure}[t]
\includegraphics[width=\columnwidth]{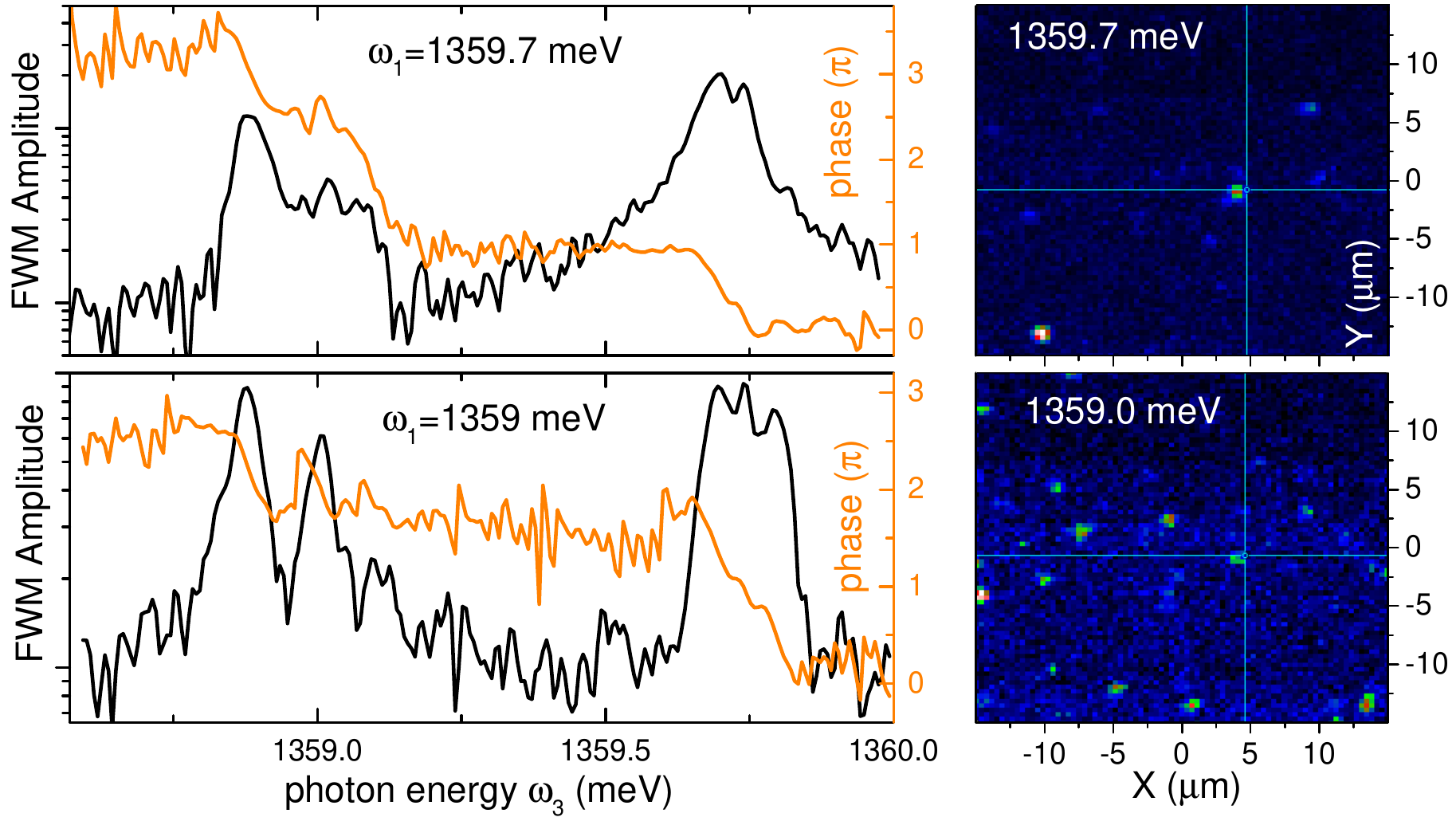}
\caption{{\bf Amplitude and phase analysis retrieved from the 2D FWM
shown in Fig.\,5 a of the main manuscript and spatial location of
coupled resonances forming a QD molecule.\,} \label{fig:FigS7}}
\end{figure}

\begin{figure}[t]
\includegraphics[width=0.5\columnwidth]{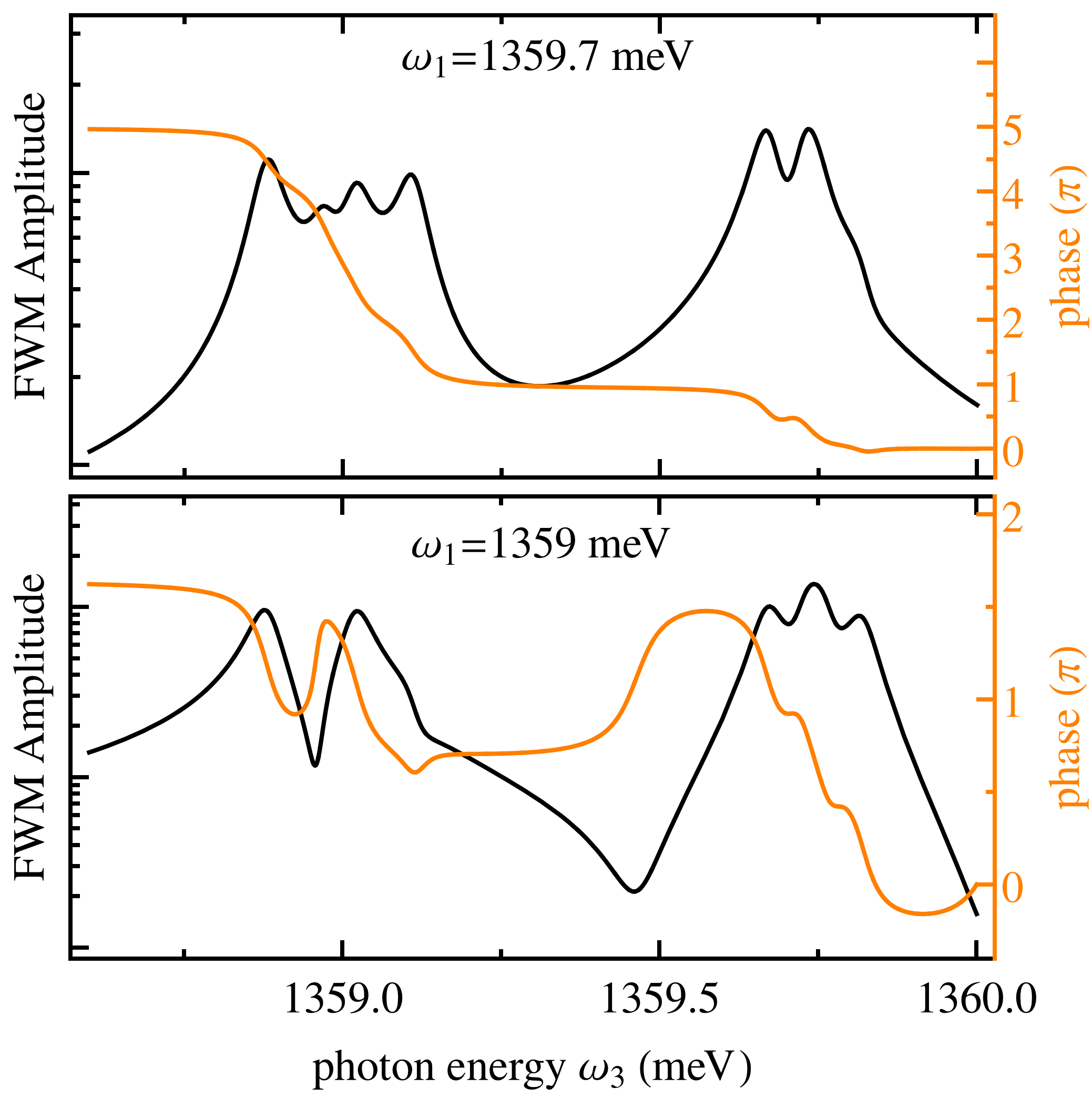}
\caption{{\bf Calculated amplitude and phase analysis, to be
compared with the measured ones shown in Fig.\,\ref{fig:FigS7}.\,}
\label{fig:FigS8}}
\end{figure}

To better illustrate coherent coupling measured at the QD molecule
investigated in Fig.\,5 of the main manuscript, in
Fig.\,\ref{fig:FigS7} we present FWM amplitude (black traces) and
phase (orange traces), retrieved from intersections through the 2D
FWM spectrum in the rephasing configuration. The diagonal peaks
display fine-structure splitting and overall phase shift of $\pi$
over the resonances. The amplitude of off-diagonals show additional
contributions, in agreement with simulations shown in
Fig\,\ref{fig:FigS8}. Each off-diagonal peak pair exhibits a
high-energy shoulder that is blueshifted by the interdot biexciton
binding energy $\Delta_{12}=90\,\mu$eV. This lifts the degeneracy of
the $G X_1$ ($G X_2$) and $X_2 B_{12}$ ($X_1 B_{12}$) contributions,
such that they do not coincide spectrally and therefore do not
cancel each other out. We also observe a 2$\pi$ phase shift across
the off-diagonals, indicating electrostatic coupling resulting in
the biexciton blueshift\,\cite{KasprzakNPho11, KasprzakJOSAB12}.
Note that the phase shifts across the off-diagonals in the
theoretical spectrum differ quantitatively from the measured ones,
since here the FSS is resolved better than in the measured spectrum.
On the other hand, the overlap with neighboring peaks has a stronger
impact in the calculated spectrum (due to the Lorentzian form of the
resonances), leading to phase interferences. As expected, both QDs
are spatially co-localized with a few hundreds of nm, as confirmed
by the FWM hyperspectral imaging. They are simultaneously excited by
the laser spot (focussed down to the diffraction limit of around
0.85$\,\mu$m) positioned at the crossing point of blue lines.

\end{document}